\documentclass[10pt,conference]{IEEEtran}
\IEEEoverridecommandlockouts
\usepackage{cite}
\usepackage{amsmath,amssymb,amsfonts}
\usepackage{algorithmic}
\usepackage{graphicx}
\usepackage{xcolor}
\usepackage{textcomp}
\usepackage{url}
\usepackage{hyperref}
\usepackage{comment}
\DeclareMathAlphabet{\mathpzc}{OT1}{pzc}{m}{it}

\begin{document}

\title{CFG2VEC: Hierarchical Graph Neural Network for Cross-Architectural Software Reverse Engineering\thanks{This material is based upon work supported by the Defense Advanced Research Projects Agency (DARPA) and Naval Information Warfare Center Pacific (NIWC Pacific) under Contract Number N66001-20-C-4024
The views, opinions, and/or findings expressed are those of the author(s) and should not be interpreted as representing the official views or policies of the Department of Defense or the U.S. Government.}
\thanks{Distribution Statement "A" (Approved for Public Release, Distribution
Unlimited).}}

\makeatletter
\newcommand{\linebreakand}{
  \end{@IEEEauthorhalign}
  \hfill\mbox{}\par
  \mbox{}\hfill\begin{@IEEEauthorhalign}
}
\makeatother

\author{
    \IEEEauthorblockN{Shih-Yuan~Yu$^1$, Yonatan~Gizachew~Achamyeleh$^1$, Chonghan~Wang$^1$, Anton~Kocheturov$^2$, Patrick Eisen$^2$,
    \linebreakand
    Mohammad~Abdullah~Al~Faruque$^1$}
    \linebreakand
    {\textit{$^1$Dept. of Electrical Engineering and Computer Science, University of California, Irvine, CA, USA}}
    \linebreakand
    {\textit{\{shihyuay, yachamye, chonghaw, alfaruqu\}@uci.edu}} 
    \linebreakand
    {\textit{$^2$Siemens Technology, Princeton, NJ, USA, \{anton.kocheturov, patrick.eisen\}@siemens.com}}
}

\maketitle

\begin{abstract}
Mission-critical embedded software is critical to our society's infrastructure but can be subject to new security vulnerabilities as technology advances.
When security issues arise, \textit{Reverse Engineers} (REs) use \textit{Software Reverse Engineering} (SRE) tools to analyze vulnerable binaries.
However, existing tools have limited support, and REs undergo a time-consuming, costly, and error-prone process that requires experience and expertise to understand the behaviors of software and vulnerabilities.
To improve these tools, we propose \textit{cfg2vec}, a Hierarchical \textit{Graph Neural Network} (GNN) based approach.
To represent binary, we propose a novel \textit{Graph-of-Graph} (GoG) representation, combining the information of control-flow and function-call graphs. 
Our \textit{cfg2vec} learns how to represent each binary function compiled from various CPU architectures, utilizing hierarchical GNN and the siamese network-based supervised learning architecture.
We evaluate \textit{cfg2vec}'s capability of predicting function names from stripped binaries.
Our results show that \textit{cfg2vec} outperforms the state-of-the-art by 24.54\% in predicting function names and can even achieve 51.84\% better given more training data.
Additionally, \textit{cfg2vec} consistently outperforms the state-of-the-art for all CPU architectures, while the baseline requires multiple training to achieve similar performance.
More importantly, our results demonstrate that our \textit{cfg2vec} could tackle binaries built from unseen CPU architectures, thus indicating that our approach can generalize the learned knowledge.
Lastly, we demonstrate its practicability by implementing it as a \textit{Ghidra} plugin used during resolving DARPA \textit{Assured MicroPatching} (AMP) challenges.
\end{abstract}

\begin{IEEEkeywords}
Software Reverse Engineering; Binary Analysis; Cross-Architecture; Machine Learning; Graph Neural Network; 
\end{IEEEkeywords}

\section{Introduction}
\label{sec:intro}

In mission-critical systems, embedded software is vital in manipulating physical processes and executing missions that could pose risks to human operators.
Recently, the \textit{Internet of Things} (IoT) has created a market valued at 19 trillion dollars and drastically grown the number of connected devices to approximately 35 billion in 2025~\cite{zhidanov2019blockchain,panarello2018blockchain,dange2020iot}.
However, while IoT brings technological growth, it unintendedly exposes mission-critical systems to novel vulnerabilities~\cite{chhetri2017cross,sargolzaei2018security, barua2022bayesimposter}.
The reported number of IoT cyberattacks increased by 300\% in 2019~\cite{landscape2019}, while the discovered software vulnerabilities rose from 1.6k to 100k~\cite{CVE-2014-0160}.
The consequence can be detrimental, as indicated in \cite{ghafoor2014analysis}, the \textit{Heartbleed} bug~\cite{heartbeat2014} can lead to a leakage of up to 64K memory, threatening not only personal but also organizational information security.
Besides, \textit{Shellshock} is a bash command-line interface shell bug, but it has existed for 30 years and remains a threat to enterprises today~\cite{shelllock2014, shelllockarticle}.
For mission-critical systems, unexpected disruptions can incur millions of dollars even if they only last for a few hours or minutes~\cite{kim2010contracting}.
As a result, timely analyzing these impacted software and patching vulnerabilities becomes critical. 

\begin{figure}[!ht]
    \centering
    \includegraphics[clip, width=.98\linewidth]{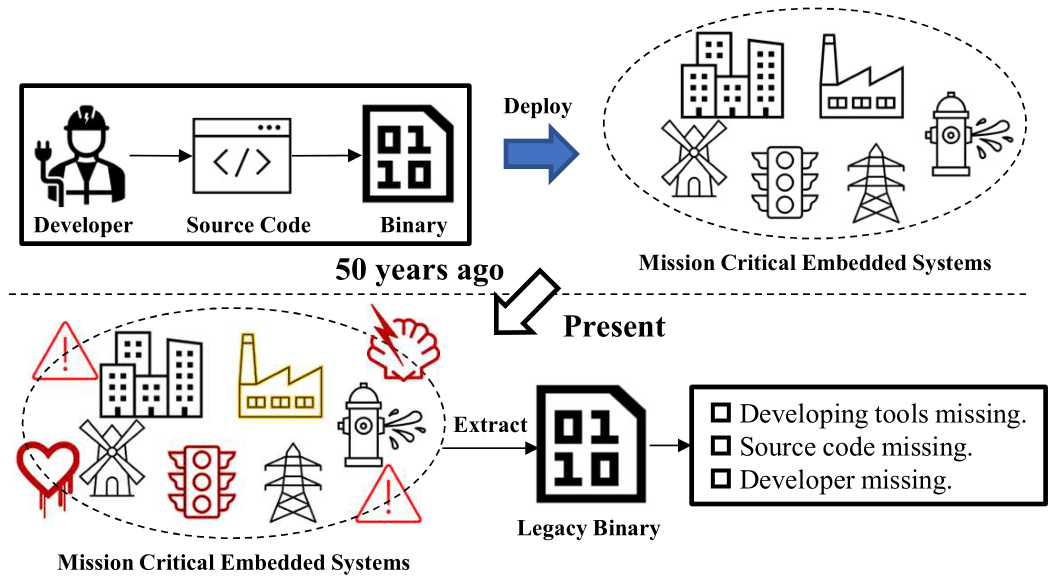}
    \caption{Legacy software life cycle.}
    \vspace{-0.5em}
    \label{fig:intro_ex}
\end{figure}

However, mission-critical systems usually use software that can last for decades due to the criticality of the missions.
Over time, these systems become legacy, and the number of newly-discovered threats can increase (as illustrated in Figure~\ref{fig:intro_ex}).
Typically, for legacy software, the original development environment, maintenance support, or source code might no longer exist.
To address vulnerabilities, vendors offer patches in the form of source code changes based on the current software version (e.g., ver 0.9).
However, the only available data in the legacy system is binary based on its source code (e.g., ver 0.1).
Such a version gap poses challenges in applying patches to the legacy binaries, leaving the only solution for applying patches to legacy software as direct binary analysis.
Today, as Figure~\ref{fig:re_flow} shows, \textit{Reverse Engineers} (REs) have to leverage \textit{Software Reverse Engineering} (SRE) tools such as \textit{Ghidra}~\cite{nsfghidra}, \textit{HexRays}~\cite{hex2017ida}, and \textit{radare2}~\cite{radare2book} to first disassemble and decompile binaries into higher-level representations (e.g., C or C++).
Typically, these tools take the debugging information, strings, and the symbol-table and binary to reconstruct function names and variable names, allowing REs to rebuild a software's structure and functionality without access to source code~\cite{keliris2018icsref}.
For REs, these symbols encode the context of the source code and provide invaluable information that could help them to understand the program's logic as they work to patch vulnerable binaries.
However, symbols are often excluded for optimizing the binary's footprint in mission-critical legacy systems where memory is limited.
Because recovering symbols from \textit{stripped binaries} is not straightforward, most decompilers assign meaningless symbol names to coding elements.
As for understanding the software semantics, REs have to leverage their experience and expertise to consume the information and then interpret the semantics of each coding element.

Recent works tackle these challenges with \textit{Machine Learning} (ML), aiming to recover the program's information from raw binaries.
For example, \cite{he2018debin}, and \cite{lacomis2019dire} associate code features to function names and model the relationships between such code features and the corresponding source-level information (variable names in \cite{lacomis2019dire}, variable \& function names in \cite{he2018debin}).
Meanwhile, \cite{david2020neural} and \cite{gao2021lightweight} use an encoder-decoder network structure to predict function names from stripped binary functions based on instruction sequences and control flows.
However, none of them support cross-architectural debug information reconstruction. 
On the other side, there exist works focusing on the cross-platform in their ML models~\cite{feng2016scalable, xu2017neural, haq2019survey}.
These works focus on modeling the binary code similarity, extracting a real-valued vector from each control-flow graph (CFG) with attributed features, and then computing the \textit{Structural Similarity} between the feature vectors of binary functions built from different CPU architectures.

In this paper, as part of a multi-industry-academia joint initiative between Siemens, the Johns Hopkins University Applied Physics Laboratory (JHU/APL), BAE Systems (BAE), and UCI, we propose \textit{cfg2vec}, which utilizes a hierarchical \textit{Graph Neural Network} (GNN) for reconstructing the name of each binary function, aiming to develop the capacity for quick patching of legacy binaries in mission-critical systems.
Our \textit{Cfg2vec} forms a \textit{Graph-of-Graph} (GoG) representation, combining CFG and FCG to model the relationship between binary functions' representation and their semantic names.
Besides, \textit{cfg2vec} can tackle cross-architectural binaries thanks to the design of Siamese-based network architecture, as shown in Figure~\ref{fig:acfg_acg}.
One crucial use case of cross-architectural decompilation is \textit{patching}, where the goal is to identify a known vulnerability or a bug and apply a patch. 
However, there can be architecture gaps when software with a bug can be compiled into many devices with diverse hardware architectures. 
For example, it is challenging to patch a stripped binary from an exotic embedded architecture compiled ten years ago that is vulnerable to a known attack such as \textit{Heartbleed}~\cite{heartbeat2014}.
While the reference patch is available in software, the reference architecture may not be readily available or documented, or the vendor may no longer exist. 
Under such circumstances, mapping code features across architectures is very helpful. 
It would allow for identifying similarities in code between a stripped binary that is vulnerable and its reference patch, even if the patch were built for a different type of CPU architecture.
For \textit{cfg2vec}, our targeted contributions are as follows:
\begin{itemize}
    \item We propose representing binary functions in \textit{Graph-of-Graph} (GoG) and demonstrate its usefulness in reconstructing function names from stripped binaries.
    \item We propose a novel methodology, \textit{cfg2vec} that uses a hierarchical \textit{Graph Neural Network} (GNN) to model control-flow and function-calling relations in binaries.
    \item We propose using cross-architectural loss when training, allowing \textit{cfg2vec} to capture the architecture-agnostic representations of binaries. 
    \item We release \textit{cfg2vec} in a GitHub repository: \url{https://github.com/AICPS/mindsight_cfg2vec}.
    \item We integrate our \textit{cfg2vec} into an experimental Ghidra plugin, assisting the realistic scenarios of patching DARPA \textit{Assured MicroPatching} (AMP) challenge binaries.
\end{itemize}

The paper is structured as follows:
Section~\ref{sec:related_work} discusses related works and fundamentals to provide a better understanding of the paper.
Section~\ref{sec:cfg2vec} describes \textit{cfg2vec}, including problem formulation, data preprocessing, and our main pipeline introduction. 
Section~\ref{sec:exp} shows our experimental results.
Lastly, we conclude the paper in Section~\ref{sec:conclusion}.
\section{Related Work}
\label{sec:related_work}
This section introduces software reverse engineering backgrounds, discusses the related works using machine learning to improve reverse engineering, and ultimately covers graph learning for binary analysis.

\begin{figure*}[!ht]
    \centering
    \includegraphics[clip, width=.98\linewidth]{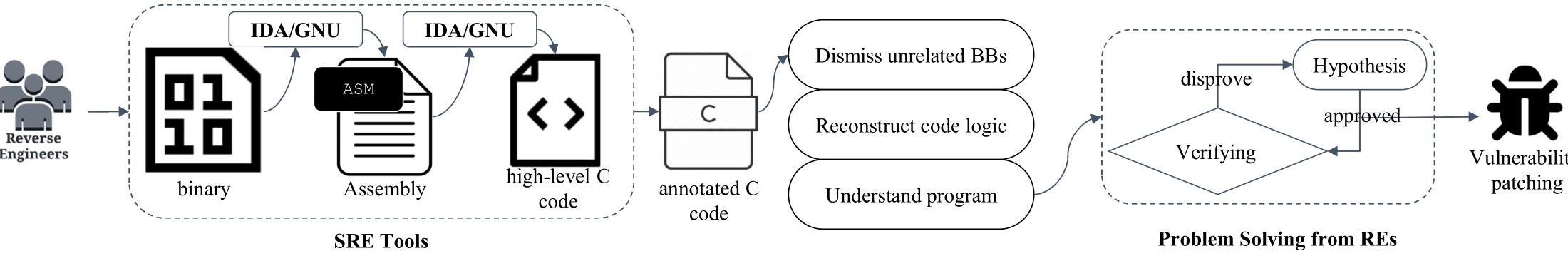}
    \caption{The RE flow to solve security issues.}
    \vspace{-1.0em}
    \label{fig:re_flow}
\end{figure*}

\subsection{Software Reverse Engineering}
\label{sec:related_work_sre}

\textit{Software Reverse Engineering} (SRE) aims at understanding the behavior of a program without having access to its source code, often being used in many applications such as detecting malware~\cite{yakdan2016helping, vdurfina2013psybot}, discovering vulnerabilities, and patching bugs in \textit{legacy software}~\cite{van2007static, brumley2013native}.
One primary tool that \textit{Reverse Engineers} (REs) use to inspect programs is \textit{disassembler} which translates a binary into low-level assembly code. 
Examples of such tools include \textit{GNU Binutils' objdump}~\cite{objdump}, \textit{IDA}~\cite{hex2017ida}, Binary Ninja~\cite{binaryninja}, and Hopper~\cite{hopper}.
However, even with these tools, reasoning at the assembly level still requires considerable cognitive effort from RE experts.

More recently, REs use \textit{decompilers} such as \textit{Hex-Rays}~\cite{hex2019}, or \textit{Ghidra}~\cite{nsfghidra} to reverse the compiling process by further translating the output of disassemblers into the code that ensembles high-level programming languages such as C or C++ to reduce the burden of understanding assembly code.
From assembly instructions, these decompilers can use program analysis and heuristics to reconstruct variables, types, functions, and control flow structure of a binary. 
However, the decompilation is incomplete even if these decompilers generate a higher-level output for better code understanding.
The reason is that the compilation process discards the source-level information and lowers its abstraction level in exchange for a smaller footprint size, faster execution time, or even security considerations.
The source-level information such as comments, variable names, function names, and idiomatic structure can be essential for understanding a program but is typically unavailable in the output of these decompilers.

As Figure~\ref{fig:re_flow} demonstrated, REs use disassemblers or decompilers to generate high-level source code.
Besides, \cite{analysisnyreyu} indicates REs will take notes and grant a name to those critical functions related to the vulnerabilities. 
This will create an annotated source code based on the high-level machine-generated source code.
While annotating the source code, REs also analyze the significant part related to the vulnerability and ignore those general instructions or unrelated codes.
At the same time, understanding the logic flow among functions is another major task they must focus on resolving their tasks.
After classification, annotation, and understanding, REs experiment with several viable remedies to find the correct patch to fix the vulnerability.

\subsection{Machine Learning for Reverse Engineering}
\label{sec:related_work_ml_re}

Software binary analysis is a straightforward first step to enhance security as developers usually deploy software in binaries~\cite{shao2022survey}.
Usually, experts conduct the patching process or vulnerability analysis by understanding the compilation source, function signatures, and variable information.
However, after the compilation, such information is usually stripped or confuscated deliberately (e.g., \textit{obfuscation}).
Software binary analysis becomes more challenging in this case as developers have to recover the source-level information based on their experience and expertise. 
The early recovery work for binaries focuses on manual completion but suffers from low efficiency, high cost, and the error-prone nature of reverse engineering.

As \textit{Machine Learning} (ML) has significantly advanced in its reasoning capability, applying ML and reconstructing higher-level source code information as an alternative to manual-based approaches has attracted considerable research attention.
For example, \cite{chuaeklavya} was the first approach that used neural network-based and graph-based models, predicting the function types to assist the reverse engineer in understanding the binary.
\cite{Aloncode2vec} also predicted function names with neural networks, aggregating the related features of sections of binary vectors.
Then, it analyzes the connections between each function in the source code (e.g., Java) and their corresponding function names for function name prediction.
\cite{he2018debin}, on the other hand, did not use a neural network.
It combined a decision-tree-based classification algorithm and a structured prediction with a probabilistic graphical model, then matched the function name by analyzing symbol names, types, and locations.
However, \cite{he2018debin} can only predict from a predetermined closed set, incapable of generalizing to new names.

As the languages for naming functions are similar to natural language, recent research works start leaning toward the use of \textit{Natural Language Processing} (NLP)\cite{artuso2019nomine, david2020neural, gao2021lightweight}.
Precisely, these models predict semantic tokens based on the function names in the library, comprising the function name during inference. 
The underlying premise is that each token corresponds in some way to the attributes and functionality of the function.
\cite{david2020neural} uses \textit{Control-Flow Graph} (CFG) to predict function names.
It combined static analysis with LSTM and transformer neural model to realize the name of functions.
However, the dataset that consisted of unbalanced data and insufficient features was limited and hindered utter performance.
\cite{artuso2019nomine} was designed to solve the limitation of the dataset.
It provided \textit{UbuntuDataset} that contained more than 9 million functions in 22K software.
\cite{gao2021lightweight} demonstrated the framework's effectiveness by building a large dataset.
It considers the fine-grained sequence and structure information of assembly code when modeling and realizing function name prediction.
Meanwhile, \cite{gao2021lightweight} reduced the diversity of data (instructions or words) while keeping the basic semantics unchanged, similar to word stemming and semantics in NLP.
However, these works have low precision scores for prediction tasks, exampled by \cite{gao2021lightweight}, only achieving around 41\% in correctly predicting the function name subtokens. 
Moreover, the metrics for the inference of unknown functions are substantially lower \cite{gao2021lightweight}, making it difficult for REs to find it helpful in practice.

Although many existing works can reconstruct source-level information, none of them supports reconstructing cross-platform debug information. 
Cross-compilation is becoming more popular in the development of software. Hardware manufacturers, for instance, often reuse the same firmware code base across several devices running on various architectures~\cite{redmond2018cross}.
A tool that performs cross-architecture function name prediction/matching would be beneficial if we have a stripped binary compiled for one architecture and a binary of a comparable program compiled for another architecture with debug symbols. We may use the binary with the debug symbols to predict the names of functions in the stripped binary, which significantly aids debugging.
A tool that could capture the architecture-agnostic characteristics of binaries would also help in malware detection as the source code of malware can be compiled in different architectures~\cite{redmond2018cross, vasan2020mthael}. Comparing two binaries of different architectures becomes more complicated because they will have different instruction sets, calling conventions, register sets, etc. Furthermore, assembly instructions from different architectures cannot often be compared directly due to the slightly different behavior of different architectures~\cite{alhanahnah2018efficient}. Cross-architecture function name prediction will assist in finding a malicious function in a program compiled for different architectures by learning its features from a binary compiled for just one architecture. The tools mentioned above are not architecture-agnostic; thus, we cannot utilize them for such applications.
To address the flaws mentioned above, aid in creating more efficient decompilers, and make reverse engineering more accessible, we propose \textit{cfg2vec}.
Incorporating the cross-architectural siamese network architecture, our \textit{cfg2vec} can learn to extract robust features that encompass platform-independent features, enhancing the state-of-the-art by achieving function name reconstruction across cross-architectural binaries. 

\subsection{Graph Learning for Binary Analysis}

\begin{figure*}[!ht]
    \centering
    \includegraphics[clip, width=.98\linewidth]{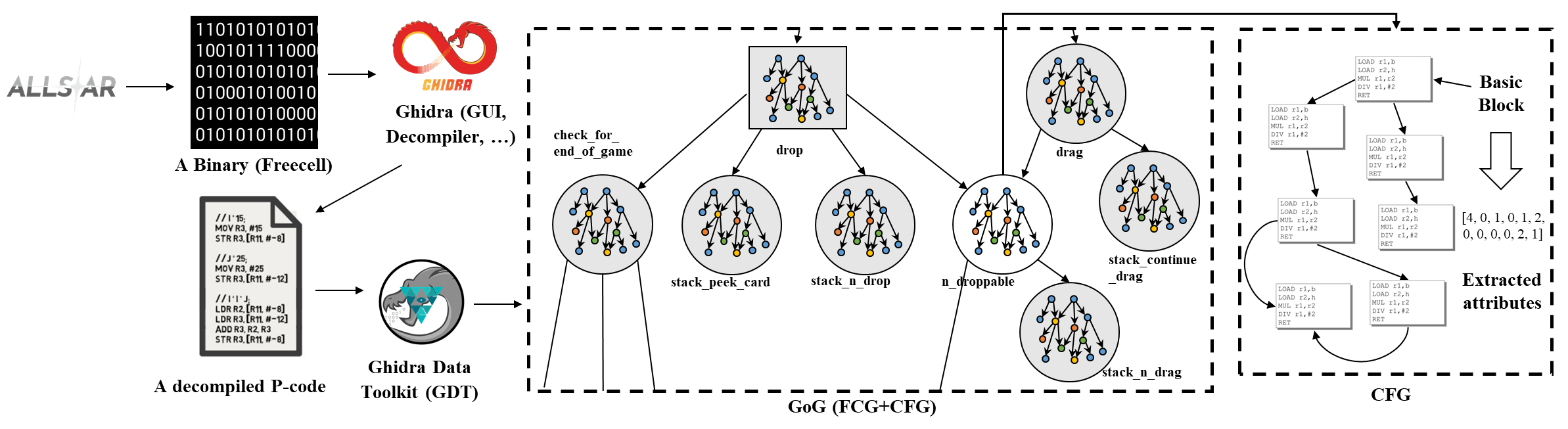}
    \caption{An example of a \textit{Graph-of-Graph} (GoG) of a binary compiled from a package Freecell with \texttt{amd64} CPU architecture.}
    \vspace{-1.5em}
    \label{fig:acfg_acg}
\end{figure*}

Graph learning has become a practical approach across fields~\cite{DBLP:journals/corr/GoyalF17, wu2020comprehensive, yasaei2021gnn4tj, yu2021scene}. 
Although conventional ML can effectively capture the features hidden in Euclidean data, such as images, text, or videos, our work focuses more on the application where the core data is graph-structured.
Graphs can be irregular, and a graph may contain a variable size of unordered nodes; moreover, nodes can have a varying number of neighboring nodes, making deep learning mathematical operations (e.g., 2D Convolution) challenging to apply.
The operations in conventional ML methods can only be applied by projecting non-Euclidean data into low-dimensional embedding space.
In graph learning, \textit{Graph Embeddings} (GE) can transform a graph into a vector (embedding of a graph) or a set of vectors (embedding of nodes or edges) while preserving the relevant and structural information about the graph~\cite{DBLP:journals/corr/GoyalF17}. 
\textit{Graph Neural Network} (GNN) is a model aiming at addressing graph-related tasks in an end-to-end manner, where the main idea is to generate a node's representation by aggregating its representation and the representations of its neighbors~\cite{wu2020comprehensive}.
GNN stacks multiple graph convolution layers, graph pooling layers, and a graph readout to generate a low-dimensional graph embedding from high-dimensional graph-structured data. 

In software binary analysis, many approaches use \textit{Control-Flow Graphs} (CFGs) as the primary representations. 
For example, \textit{Genius} forms an \textit{Attributed Control-Flow Graph} (ACFG) representation for each binary function by extracting the raw attributes from each \textit{Basic Block} (BB), a straight-line code sequence with no branching in or out except at the entry and exit, in an ACFG~\cite{feng2016scalable}.
\textit{Genius} measures the similarity of a pair of ACFGs through a bipartite graph matching algorithm, and the ACFGs are then clustered based on similarity.
\textit{Genius} leverages a codebook for retrieving the embedding of an ACFG based on similarity. 
Another approach, \textit{Gemini}, proposes a deep neural network-based model along with a Siamese architecture for modeling binary similarities with greater efficiency and accuracy than other state-of-the-art models of the time~\cite{xu2017neural}.
\textit{Gemini} takes in a pair of ACFGs extracted from raw binary functions generated from known vulnerability in code and then embeds them with a shared \textit{Structure2vec} model in their network architecture. Once embedded, \textit{Gemini} trains its model with a loss function that calculates the cosine similarities between two embedded representations.
\textit{Gemini} outperforms models like \textit{Genius} or other approaches such as bipartite graph matching.
In literature, there exist other works that consider the \textit{Function Call Graph} (FCG) as their primary data structures in binary analysis for malware detection~\cite{hassen2017scalable}.
Our \textit{cfg2vec} extracts relevant platform-independent features by combining the usage of CFG and FCG, resulting in a \textit{Graph-of-Graph} (GoG) representation for cross-architectural high-level information reconstruction tasks (e.g., function name).
\section{CFG2VEC Architecture}
\label{sec:cfg2vec}

This section begins with problem formulation. 
Next, as Figure~\ref{fig:archi} shows, we depict how our \textit{cfg2vec} extracts the \textit{Graph-of-Graph} (GoG) representation from each software binary. 
Lastly, we describe the network architecture in \textit{cfg2vec}.

\subsection{Problem Formulation}
In our work, given a binary code, denoted as $p$, compiled from different CPU architectures, we extract a graph-of-graph (GoG) representation, $\mathpzc{G} = (\mathpzc{V}, \mathpzc{A})$ where $\mathpzc{V}$ is the set of nodes and $\mathpzc{A}$ is the adjacency matrix (As Figure~\ref{fig:acfg_acg} shows). The nodes in $\mathpzc{V}$ represent functions and the edges in $\mathpzc{A}$ indicate their cross-referencing relationships.
That says, each of the node $f_i \in \mathpzc{V}$ is a CFG, and we denote it as $f_i = (B, A, \phi)$ where the nodes in $B$ represent the basic blocks and the edges in $A$ denote their dependency relationships. $\phi$ is a mapping function that maps each basic block in the assembly form to its corresponding extracted attributes $\phi(v_i) = C^k$ where C is a numeric value, and k is the number of attributes for the basic block (BB).
Whereas the CFG structure is meant to provide more information at the lower BB level, the GoG structure is intended for recovering information at the overarching function level between the CFGs.
Figure \ref{fig:acfg_acg} is an example of a partial GoG structure with a closer inspection of one of its CFG nodes and another of a single CFG BB node, showing the set of features corresponding to that BB node. 
The goal is to design an efficient and effective graph embedding technique that can be used for reconstructing the function names for each function $f_i \in \mathpzc{V}$.

\subsection{Ghidra Data ToolKit for Graph Extraction}
\label{sec:gdt}
To extract the structured representation required for \textit{cfg2vec} we leverage the state-of-the-art decompiler \textit{Ghidra}~\cite{nsfghidra} and the \textit{Ghidra Headless Analyzer}\footnote{Documentation of \textit{Ghidra Headless Analyzer}: \url{https://ghidra.re/ghidra_docs/analyzeHeadlessREADME.html}}.
The \textit{headless analyzer} is a command-line version of \textit{Ghidra} allowing users to perform many tasks (such as analyzing a binary file) supported by \textit{Ghidra} via a command-line interface.
For extracting GoG from a binary, we developed our \textit{Ghidra Data Toolkit} (GDT);
GDT is a set of Java-based metadata extraction scripts used for instrumenting \textit{Ghidra Headless Analyzer}.
First, GDT programmatically analyzes the given executable file and stores the extracted information in the internal Ghidra database. 
Ghidra provides a set of APIs to access the database and retrieve the information about the analyzed binary.
GDT uses these APIs to export information such as Ghirda's PCode and call graph for each function.
Specifically, the \textit{FunctionManager} API  allows us to manipulate the information of each decompiled function in the binary and acquire the cross-calling dependencies between functions.
For each function, we utilized another Ghidra API \textit{DecompInterface}\footnote{Documentation of \textit{Ghidra API DecompInterface}: \url{https://ghidra.re/ghidra_docs/api/ghidra/app/decompiler/DecompInterface.html}} to extract 12 attributes associated with each basic block in a function.
These attributes precisely correspond to the total number of instructions, including arithmetic, logic, transfer, call, data transfer, SSA, compare, and pointer instructions, as well as other instructions not falling within those categories and  the total number of constants and strings within that BB. 
Lastly, by integrating all of the information, we form a GoG representation $\mathpzc{G}$ for each binary $p$. 
We repeat this process until all binaries are converted to the GoG structure. We feed the resulting GoG representations to our model in batches, with the batch size denoted as B.

\subsection{Hierarchical Graph Neural Network}
\label{sec:cfg2vec_approach}

\begin{figure*}[!ht]
    \centering
    \includegraphics[width=\linewidth]{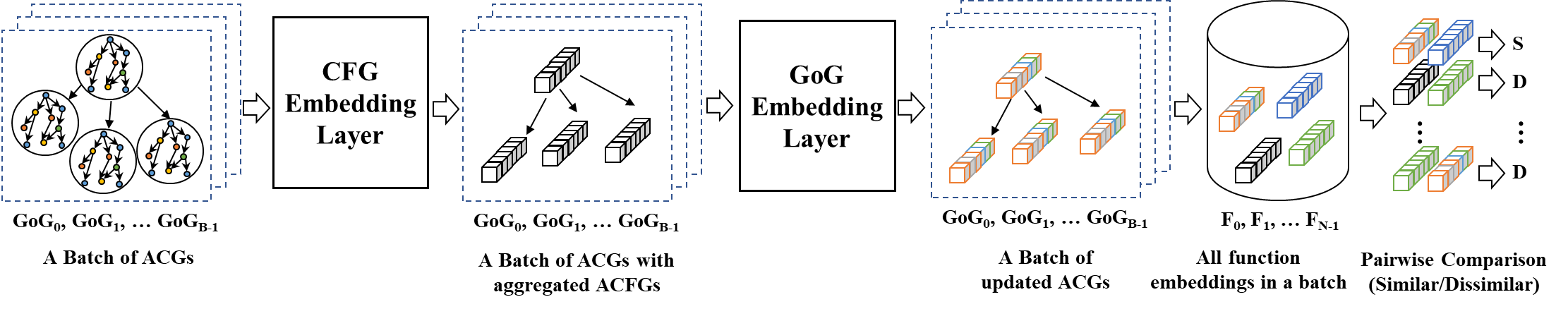}
    \vspace{-1em}
    \caption{The architecture of \textit{cfg2vec} with a supervised hierarchical graph neural network approach.}
    \vspace{-1.0em}
    \label{fig:archi}
\end{figure*}

Once $\mathpzc{G}$ is extracted from the GDT, we then feed it to our hierarchical network architecture (inspired from~\cite{harada2020dual}) that contains both \textit{CFG Graph Embedding} layer and 
\textit{GoG Graph Embedding Layer} as Figure~\ref{fig:archi} shows.
For each GoG structure, we denote it as $\mathpzc{G} = (\mathpzc{V}, \mathpzc{A})$ where $\mathpzc{V}$ is a set of functions associated with $\mathpzc{G}$ and $\mathpzc{A}$ indicates the calling relationships between the functions in $\mathpzc{V}$. Each function in $\mathpzc{V}$ is in the form of CFG $f_i = (B, A, \phi)$ where each node $b\in B$ is a BB represented in a fixed-length attributed vector $b\in R^d$, and $d$ is the dimension that we have mentioned earlier. $A$ encodes the pair-wise control-flow dependency relationships between these BBs.

\subsubsection{CFG Graph Embedding Layer} Our network architecture first feeds all functions in a batch of GoGs to the \textit{CFG Graph Embedding Layer} consisting of multiple graph convolutional layers and a graph readout operations.
The input to this layer is a function $f_i = (B, A, \phi)$ and the output is the fixed-dimensional vector representing a function. For each BB $b_k$ we let $b_k^0 = b_k$, and we update $b_k^t$ to $b_k^{t+1}$ with the graph convolution operation shown as follows:
\begin{equation*}
    b_k^{t+1} = f_G (Wb_k^t + \sum_{b_m \in A_k} Mb_m^t)
\end{equation*}
where $f_G$ is a non-linear activation function such as ReLU, $A_k$ is the list of adjacent BBs for $b_k$, and $W\in R^{d\times d}$ and $M\in R^{d\times d}$ are the weights to be learned during the training. We run $T$ iterations of such a convolution, which can be a tunable hyperparameter in our model. During the updates, each BB gradually aggregates the global information of the control-flow dependency relations into its representation, utilizing the representation of its neighbor. We obtain the final representation for each BB as $b_k^{T}$. To acquire the representation for the function $f_i$, we apply a graph readout operation such as \textit{sum-readout}, described as follows,
\begin{equation}
    g^{(T)} = \sum_{b_k \in B} b_k^{T}
\end{equation}
We assign the value of $g^{(T)}$ (a.k.a. CFG embedding) to $f_i$. The graph readout operation can be replaced with \textit{mean-readout} or \textit{max-readout}.

\subsubsection{GoG Graph Embedding Layer} Once all the functions have been converted to fixed-length graph embeddings, we then feed $\mathpzc{G}$ to the second layer of \textit{cfg2vec}, the \textit{GoG Embedding Layer}. Here, for each function $f_i$ we apply another $L$ iterations of graph convolution with $\mathpzc{F}$ and $\mathpzc{C}$. The updates can be illustrated as follows, 
\begin{equation}
    f_k^{(l+1)} = f_{GoG} (Uf_k^l + \sum_{f_m \in C_k} Vf_m^{(l)})
\end{equation}
where $f_{GoG}$ is a non-linear activation function and $C_k$ is the list of adjacent functions (calling) for the function $f_k$ and $U\in R^{d\times d}$ and $V\in R^{d\times d}$ are the weights to be learned during the training. Lastly, we take the $f_k^{(L)}$ as the representation that considers both CFG and GoG graph structures. We use these updated representations to perform cross-architecture function similarity learning.

\subsubsection{Siamese-based Cross-Architectural Function Similarity}
Given a batch of GoGs $B=\{GoG_1, GoG_2, ..., GoG_B\}$, we apply the hierarchical graph neural network to acquire the set of updated function embeddings, denoted as $B_F = \{f_1^{(T)}, f_2^{(T)}, ..., f_K^{(T)}\}$. We calculate the function similarity for each function pair with cosine similarity, denoted as $\hat{y} \in [-1, 1]$. The loss function $J$ between $\hat{Y}$ and a ground-truth label $y$, which indicates whether a pair of functions have the same function or not, is calculated as follows, 
\begin{equation}
    J (\hat{y}, y) = \left \{
    \begin{array}{ll}
         1-y, &  \text{if y=1,}\\
         MAX(0, \hat{y}-m), &  \text{if y=-1,}
    \end{array}
    \right.
\end{equation}
the final loss $L$ is then calculated as follows,
\begin{equation}
    L = H (Y, \hat{Y}) = \sum_i(J(\hat{y_i}, y_i)),
\end{equation}
where $Y$ stands for ground-truth labels (either similarity or dissimilarity), and $\hat{Y}$ represents the corresponding predictions. More specifically, we denote a pair of functions as similar if they are the same but compiled with different CPU architectures. The $m$ is a constant to prevent the learned embeddings from becoming distorted (by default, $0.5$). 
To maintain the balance between positive and negative training samples, we developed a custom batching algorithm. 
The function leverages the knowledge gained by adding a binary of some package to a given batch to find and add a binary for the same package, built for a different architecture, to the provided batch as a positive sample. It will also include a binary from another package as a negative sample. This will give any batch a balanced proportion of positive and negative samples.
Finally, we use the loss $L$ to update all the associated weights in our neural networks with an \textit{Adam} optimizer. 
Once trained, we then use the model to perform function name reconstruction tasks.
\section{Experimental Results}
\label{sec:exp}
In this section, we evaluate \textit{cfg2vec}'s capability in predicting function names.
We first describe the dataset preparation and the training setup processes.
Then, we present the comparison of \textit{cfg2vec} against baseline in predicting function names. 
Although many baseline candidates tackle the same problem~\cite{he2018debin, artuso2019nomine, david2020neural, gao2021lightweight}, some require purchasing a paid version of IDA Pro to preprocess datasets, and some even do not open source their implementations. 
Therefore, \cite{he2018debin} was the only feasible choice, as running other models using our datasets was almost impossible.
Next, we also show the result of the ablation study over  \textit{cfg2vec}.
Besides, we exhibit that our \textit{cfg2vec} can perform architecture-agnostic prediction better than the baseline.
Lastly, we illustrate the real-world use case where our \textit{cfg2vec} is integrated as a \textit{Ghidra} plugin application for assisting in resolving challenging reverse engineering tasks.
We conducted all experiments on a server equipped with Intel Core i7-7820X CPU @3.60GHz with 16GB RAM and two NVIDIA GeForce GTX Titan Xp GPUs.

\subsection{Dataset Preparation}
\label{sec:data_prep}
Our evaluating data source is the ALLSTAR (\textit{Assembled Labeled Library for Static Analysis Research}) dataset, hosted by \textit{Applied Physics Laboratory} (APL)~\cite{allstarDataset}.
It has over 30,000 Debian Jessie packages pre-built from \texttt{i386}, \texttt{amd64}, \texttt{ARM}, \texttt{MIPS}, \texttt{PPC}, and \texttt{s390x} CPU architectures for software reverse engineering research.
The authors used a modified \textit{Dockcross} script in docker to build each package for each supported architecture. 
Then, they saved each resulting ELF with its symbols, the corresponding source code, header files, intermediate files (.o, .class, .gkd, .gimple), system headers, and system libraries altogether.

To form our datasets, we selected the packages that have ELF binaries built for the \texttt{amd64}, \texttt{armel}, \texttt{i386}, and \texttt{mipsel} CPU architectures.
\texttt{i386} and \texttt{amd64} are widely used by general computers, especially in the Intel and AMD products, respectively. 
\texttt{MIPS} and \texttt{ARM} are crucial in embedded systems, smartphones, and other portable electronic devices~\cite{Debian-GNU-Linux}.
In practice, we excluded the packages with only one CPU architecture in the ALLSTAR dataset.
Additionally, due to our limited local computing resources, we eliminated packages that were too large to handle.
We checked each selected binary on whether the ground-truth symbol information exists using the \textit{Ghidra} decompiler and Linux \texttt{file} command and removed the ones that do not have them.
Lastly, we assembled our primary dataset, called the \textit{AS-4cpu-30k-bin} dataset, that consists of 27572 pre-built binaries from 1117 packages and 4 CPU architectures, as illustrated in Table~\ref{tab:data_stat}.

Our preliminary experiment revealed that the evaluation had a data leakage issue when splitting the dataset randomly.
Therefore, we performed a non-random variant train-test split with a 4-to-1 ratio on the \textit{AS-4cpu-30k-bin} dataset, selecting roughly 80\% of the binaries for the training dataset and leaving the rest for the testing dataset.
We referenced \cite{xu2017neural} for their splitting methods, aiming to ensure that the binaries that belong to the same packages stay in the same set, either the training or testing sets.
Such a variant splitting method allows us to evaluate \textit{cfg2vec} truly.

Next, we converted binaries in the \textit{AS-4cpu-30k-bin} dataset into their \textit{Graph-of-Graph} (GoG) representations leveraging the GDT mentioned previously in Section~\ref{sec:gdt}.
Notably, we processed a batch of binaries related to one package at one time as developers might define user functions in different modules of the same package while putting prototype declarations in that package's main module.
For this case, \textit{Ghidra} indeed recognizes two function instances while one only contains the function declaration and another has its actual function content.
As these two instances correspond to the same function name and one contains only dummy instructions, they can thus create noise in our datasets, thus affecting our model's learning.
To cope with this, our GDT also searches from other binaries of the same package for the function bodies.
If found, our GDT associates that user function with the function graph node with the actual content data.
Besides user functions, library function calls may exist, and searching their function bodies in the same package would fail for dynamically loaded binaries.
Under such circumstances, \textit{Ghidra} would recognize these functions as \textit{ThunkFunctions}\footnote{ThunkFunction Manual: \url{https://ghidra.re/ghidra_docs/api/ghidra/program/model/listing/ThunkFunction.html}} which only contain one dummy instruction.
As a workaround, we removed these \textit{ThunkFunctions} from our data as they might mislead the model's learning.
Applying this workaround indicates that our model works in predicting function names for the user and statically linked functions.

\begin{table}[htbp]
    \centering
    \caption{The statistics of datasets used in our experiments.}
    \vspace{-0.5em}
    \begin{tabular}{ p{2.0cm} | c c c c c}
        \hline
        Dataset / \#& pkg/bin & func node/edge$^{\mathrm{1}}$ & bb node/edge$^{\mathrm{2}}$ \\
        \hline
        \textit{AS-4cpu-30k-bin} & 1117/27,572 & 51.17/97.14 & 14.12/19.98 \\
        \hline
        \textit{AS-3cpu-9k-bin}& 633/9,000 & 44.01/79.06 & 12.24/17.07 \\
        \hline
        \textit{AS-i386-3k-bin}& 633/3,000 & 45.31/87.70 & 11.45/15.97 \\
        \hline
        \textit{AS-amd64-3k-bin} & 633/3,000 & 42.28/74.07 & 12.28/17.18 \\
        \hline
        \textit{AS-armel-3k-bin} & 633/3,000 & 44.45/75.41 & 13.00/18.07 \\
        \hline
        \multicolumn{4}{l}{$^{\mathrm{1}}$ \# of average functions and edges in each binary}\\
        \multicolumn{4}{l}{$^{\mathrm{2}}$ \# of average bb blocks and edges from each function}
    \end{tabular}
    \label{tab:data_stat}
\end{table}

\begin{table*}[!ht]
    \centering
    \caption{The performance evaluation of \textit{cfg2vec} for function name prediction against \cite{he2018debin}.}
    \vspace{-0.5em}
    \begin{tabular}{ p{1.8cm} | c | c | c c c c c }
        \hline
        Model & Training dataset & Testing dataset & P@1$^{\mathrm{1}}$ & P@2$^{\mathrm{1}}$ & P@3$^{\mathrm{1}}$ & P@4$^{\mathrm{1}}$ & P@5$^{\mathrm{1}}$\\
        \hline
        \textit{cfg2vec} & \textit{AS-4cpu-30k-bin} & \textit{AS-noMipsel-300-bin} & 97.05\% & 99.47\% & 99.47\% & 99.47\% & 99.47\% \\
        \hline
        \textit{cfg2vec} & \textit{AS-4cpu-20k-bin} & \textit{AS-noMipsel-300-bin} & 74.22\% & 75.76\% & 75.78\% & 75.78\% & 78.78\% \\
        \hline
        &  & \textit{AS-amd-100-bin} & 69.18\% & 69.98\% & 69.98\% & 69.98\% & 69.98\% \\
        \textit{cfg2vec} &  \textit{AS-3cpu-9k-bin} & \textit{AS-i386-100-bin} & 69.41\% & 70.39\% & 70.39\% & 70.39\% & 70.39\% \\
        & & \textit{AS-armel-100-bin} & 70.66\% & 71.04\% & 71.11\% & 71.11\% & 71.11\% \\
        & & \textit{AS-noMipsel-300-bin} & 69.75\% & 70.47\% & 70.50\% & 70.50\% & 70.50\% \\

        \hline
        \cite{he2018debin}-\texttt{amd64}$^{\mathrm{2}}$ &  \textit{AS-amd64-3k-bin} & \textit{AS-amd-100-bin} & 29.32\% & - & - & - & -\\
        \hline
        \cite{he2018debin}-\texttt{i386}$^{\mathrm{2}}$ &  \textit{AS-i386-3k-bin} & \textit{AS-i386-100-bin} & 52.64\% & - & - & -& -\\
        \hline
        \cite{he2018debin}-\texttt{armel}$^{\mathrm{2}}$ & \textit{AS-armel-3k-bin} & \textit{AS-armel-100-bin} & 53.65\% & - & - & - & -\\
        \hline
        \multicolumn{8}{l}{$^{\mathrm{1}}$ P@k measures if the actual function name is in the top k of the predicted function names.}\\
        \multicolumn{8}{l}{$^{\mathrm{2}}$ These models only provide the top 1 function name prediction; hence they only have P@1 value.}\\
    \end{tabular}
    \vspace{-1.0em}
    \label{tab:cfg2vecfn}
\end{table*}

We experimented \cite{he2018debin} with our datasets, referencing to their implementation\footnote{Debin's~\cite{he2018debin} repository: \url{https://github.com/eth-sri/debin}}.
As \cite{he2018debin} used a dataset with 3,000 binaries for experiments, we followed accordingly, preparing datasets with smaller but similar sizes.
We achieved this by downsampling from our primary \textit{AS-4cpu-30k-bin} dataset, creating the \textit{AS-3cpu-9k-bin} dataset which has 9,000 binaries for \texttt{i386}, \texttt{amd64}, and \texttt{armel} CPU architectures.
Furthermore, as \cite{he2018debin} supports only one CPU architecture at a time, we then separated the \textit{AS-3cpu-9k-bin} dataset into different CPU architectures, generating three training datasets for testing \cite{he2018debin}: \textit{AS-i386-3k-bin}, \textit{AS-amd64-3k-bin}, and \textit{AS-armel-3k-bin}.
For training, we utilized the \texttt{strip} Linux command, converting our original data into three: the original binaries (\textit{debug}), stripped binaries with debug information (\textit{stripped}), and stripped binaries without debug information (\textit{stripped\_wo\_symtab}) to follow \cite{he2018debin}'s required data format.
For evaluation, we sampled 100 binaries from our primary dataset for each CPU architecture, labeled \textit{AS-amd-100-bin}, \textit{AS-i386-100-bin}, \textit{AS-armel-100-bin}, and \textit{AS-mipsel-100-bin}. 
We also have another evaluation dataset called \textit{AS-noMipsel-300-bin}, which contains roughly 300 binaries produced for the \texttt{amd64}, \texttt{i386}, and \texttt{armel} platforms.
Table~\ref{tab:data_stat} summarizes the data statistics for all these datasets, including the numbers of packages and binaries, the average number of function nodes, edges, and BB nodes.
The following sections will detail how we utilized these datasets during our experiments.

\subsection{Evaluation: Function Name Prediction}
\label{sec:exp_pred}

Table~\ref{tab:cfg2vecfn} demonstrates the results of  \textit{cfg2vec} in predicting function names.
For the baseline, we followed \cite{he2018debin}'s best setting where the feature dimension of register or stack offset are both 100 to train with our prepared datasets.
For \textit{cfg2vec}, we used three GCN layers and one GAT convolution layer in both graph embedding layers.
For evaluation, we calculate the p@k (e.g., precision at k) metric, which refers to an average hit ratio over the top-k list of predicted function names. 
Specifically, we feed each binary represented in GoG into our trained model, converting each function $f \in F$ and acquiring its function embedding $h_f$.
Then, we calculate pair-wise cosine similarities between $h_f$ and all the other function embeddings, forming a top-k list by selecting k names in which their embeddings are top-kth similar to $h_f$.
If the ground-truth function name is among the top-k list of function name predictions, we regard that as a hit; otherwise, it is a miss.
During experiments, we set the top-k value to be 5, so our model can recommend the best five possible names for each function in a binary.

As shown in Table~\ref{tab:cfg2vecfn}, \textit{cfg2vec}, trained with the \textit{AS-3cpu-9k-bin} dataset, can achieve a 69.75\% prediction accuracy (e.g., p@1) in inferring function names.
For \cite{he2018debin}, we had to train their models for each CPU architecture separately as it cannot train in a cross-architectural manner.
Even so, for \texttt{amd64} binaries, \cite{he2018debin} only achieves 29.32\% precision, while for \texttt{i386} and \texttt{armel}, it performs 52.64\% and 53.65\%, respectively.
This result indicates that in any case, our \textit{cfg2vec} outperforms \cite{he2018debin}.
Besides, while \cite{he2018debin} only yields one prediction, our \textit{cfg2vec} suggests five choices, making it flexible for our users (e.g., REs) to select what they believe best fits the function among the best k predicted names.
The p@2 to p@5 in Table~\ref{tab:cfg2vecfn} demonstrate that our \textit{cfg2vec} can  provide enough hints of function names for users. 
For example, p@5 of \textit{cfg2vec} trained with our \textit{AS-3cpu-9k-bin} dataset can achieve 70.50\% precision across all the CPU architecture binaries.
We also experimented our \textit{cfg2vec} with larger datasets.
From Table~\ref{tab:cfg2vecfn}, we can observe that \textit{cfg2vec} can have 5.04\% performance gain in correctly predicting function names (e.g., p@1). 
Moreover, the gain increases to 28\% when training \textit{cfg2vec} with the \textit{AS-4cpu-30k-bin} dataset.
We believe training on a larger dataset implies training with a more diversified set of binaries.
This allows our model to acquire more knowledge, thus being capable of extracting more robust features for binary functions.
In summary, this result indicates that compared to the baseline, our model can effectively provide contextually relevant names for functions in the decompiled code to our users.

\begin{table}[!ht]
    \centering
    \caption{The comparison between \textit{cfg2vec} and its ablated variations.}
    \vspace{-0.5em}
    \begin{tabular}{c | c | c| c | c | c }
        \hline
         Arch & \cite{he2018debin} &  \textit{GCN-GAT} & \textit{2GCN} &   \textit{2GCN-GAT} & \textit{cfg2vec}   \\
          \hline
        \texttt{amd64} & 29.32\%$^{\mathrm{1}}$ & 61.59\% & 69.49\% & 69.56\% & 70.66\%  \\
        \hline
        \texttt{armel} & 52.64\%$^{\mathrm{2}}$  & 66.40\% & 68.59\% & 68.92\% & 69.19\%\\
        \hline
        \texttt{mipsel} & 53.65\%$^{\mathrm{3}}$   & 66.47\% & 68.17\% & 68.56\% & 69.41\%\\
        \hline
        \hline
        Overall & 45.20\% & 64.82\% & 68.75\% & 69.01\% & 69.75\%\\
        \hline
        \multicolumn{6}{l}{$^{\mathrm{1}}$ Evaluation results for \cite{he2018debin}-\texttt{amd64} model.}\\
        \multicolumn{6}{l}{$^{\mathrm{2}}$ Evaluation results for \cite{he2018debin}-\texttt{i386} model.}\\
        \multicolumn{6}{l}{$^{\mathrm{3}}$ Evaluation results for\cite{he2018debin}-\texttt{armel} model.}
    \end{tabular}
    \label{tab:ablation}
\end{table}

We also experimented with various ablated network setups to study how each component of \textit{cfg2vec} contributes to performance. 
First, we simplified our \textit{cfg2vec} by stripping one GCN layer from the original experimental setup.
As shown in Table~\ref{tab:ablation}, we called this setup \textit{2GCN-GAT} which slightly decreased the performance by 0.75\%.
Then, from \textit{2GCN-GAT} setup, we further removed the GAT layer, calling it \textit{2GCN}. 
We again observed a marginal performance decrease ($<$1\%).
Next, we eliminated another GCN layer from \textit{2GCN-GAT}, constructing the \textit{GCN-GAT} setup. 
For \textit{GCN-GAT}, we saw a drastic drop (4.2\%) which highlights that the number of GCN layers can be an essential factor in the performance.
Specifically, we found that going from 1 to 2 GCN layers improves prediction accuracy by more than 4\%. 
However, we do not observe a significant performance gain when increasing the number of GCN layers to more than three. 
Therefore, we retained the original \textit{cfg2vec} model with its three GCN layers.
All in all, as shown in Table~\ref{tab:ablation}, all these ablated models, still outperform  \cite{he2018debin}, which we attributed to the GoG representation we made for each binary in the dataset.


\subsection{Evaluation: Architectural-agnostic Prediction}
\label{sec:exp_cross}

Table~\ref{tab:crossarchi} demonstrates our \textit{cfg2vec}'s capability in terms of cross-architecture support.
As \cite{he2018debin} supports training one CPU architecture at a time, we had to train it multiple times during experiments.
Specifically, we trained \cite{he2018debin} on three datasets: \textit{AS-amd64-3k-bin}, \textit{AS-i386-3k-bin}, and \textit{AS-armel-3k-bin}, calling resulting trained models, \cite{he2018debin}-\texttt{amd64}, \cite{he2018debin}-\texttt{i386}, and \cite{he2018debin}-\texttt{armel}, respectively.
For these baseline models, we observe that they perform well when tested with the binaries built on the same CPU architecture but poorly with the ones built on different CPU architectures.
For instance, \cite{he2018debin}-\texttt{amd64} achieves 29.3\% accuracy for \texttt{amd64} binaries, but performs worse for \texttt{i386} and \texttt{armel} binaries (13.8\% and 7.1\%).
Similarly, \cite{he2018debin}-\texttt{i386} achieves 52.6\% accuracy for \texttt{i386} binaries, but performs worse for \texttt{amd64} and \texttt{armel} binaries (6.2\% and 1.1\%).
Lastly, \cite{he2018debin}-\texttt{armel} achieves 53.6\% accuracy for \texttt{armel} binaries, but performs worse for \texttt{amd64} and \texttt{i386} binaries (11.8\% and 8.9\%).
We used the top-1 prediction generated from \textit{cfg2vec} (a.k.a., p@1) as the comparing metric as \cite{he2018debin} produces only one prediction per each function.
From the results, we observe that \textit{cfg2vec} outperforms \cite{he2018debin} across all three tested CPU architectures.
The fact that \textit{cfg2vec} performs consistently well across all CPU architectures indicates that our \textit{cfg2vec} supports cross-architecture prediction. 

To evaluate the capability of generalizing the learned knowledge, we tested all models with the \textit{AS-mipsel-100-bin} dataset, which has binaries built from another famous CPU architecture, \texttt{mipsel}, that our \textit{cfg2vec} does not train before.
For \cite{he2018debin}, it has lower performance when testing on binaries built from the CPU architectures that it did not train before, exampled by the highest accuracy of \cite{he2018debin} to be 13.84\% when trained on \textit{amd64} binaries and evaluated on \texttt{i386} binaries. 
In our work, as Table~\ref{tab:crossarchi} shows, our \textit{cfg2vec} achieves 36.69\% accuracy when trained with \texttt{amd64}, \texttt{i386}, and \texttt{armel} binaries but tested on \texttt{mipsel} binaries.
For \cite{he2018debin}, it does not even support analyzing \texttt{mipsel} binaries.
In short, these results demonstrate that our \textit{cfg2vec} outperforms our baseline in the function name prediction task on cross-architectural binaries and generalizes better to the binaries built from unseen CPU architectures.
To further investigate \textit{cfg2vec}'s cross-architecture performance, we trained it on three datasets, each consisting of binaries built for two different architectures. 
We then gave the resulting trained models names that indicated the architectures from which the binaries were derived: \textit{cfg2vec}-armel-i386, \textit{cfg2vec}-amd64-i386, and \textit{cfg2vec}-armel-amd64. 
These results show that our model performs well in the function name prediction job across all of these scenarios, including when tested on binaries compiled for unknown CPU architectures.

\begin{table}[!ht]
    \centering
    \caption{The cross-architectural comparison between cfg2vec and \cite{he2018debin}}
    \vspace{-0.5em}
    \begin{tabular}{ |p{2.5cm} | p{2.3cm} | r |}
        \hline
        Model & Testing dataset & P@1  \\\hline
        & \textit{AS-amd-100-bin} & 69.18\%  \\
        \textit{cfg2vec}-3-Arch & \textit{AS-i386-100-bin} & 70.66\% \\
        & \textit{AS-armel-100-bin} & 69.41\%  \\
        & \textit{AS-mipsel-100-bin}\textcolor{blue}{$^{\mathrm{*}}$} & 36.69\% \\
        \hline
        & \textit{AS-amd-100-bin} & 68.53\%  \\
        \textit{cfg2vec}-amd64-armel & \textit{AS-i386-100-bin}\textcolor{blue}{$^{\mathrm{*}}$} & 39.23\% \\
        & \textit{AS-armel-100-bin} & 69.11\%  \\
        & \textit{AS-mipsel-100-bin}\textcolor{blue}{$^{\mathrm{*}}$} & 32.21\% \\
        \hline
        & \textit{AS-amd-100-bin} & 68.59\%  \\
        \textit{cfg2vec}-amd64-i386 & \textit{AS-i386-100-bin} & 69.09\% \\
        & \textit{AS-armel-100-bin}\textcolor{blue}{$^{\mathrm{*}}$} & 34.20\%  \\
        & \textit{AS-mipsel-100-bin}\textcolor{blue}{$^{\mathrm{*}}$} & 38.26\% \\
        \hline
        & \textit{AS-amd-100-bin}\textcolor{blue}{$^{\mathrm{*}}$} & 42.96\%  \\
        \textit{cfg2vec}-armel-i386 & \textit{AS-i386-100-bin} & 67.45\% \\
        & \textit{AS-armel-100-bin} & 63.86\%  \\
        & \textit{AS-mipsel-100-bin}\textcolor{blue}{$^{\mathrm{*}}$} & 36.61\% \\
        \hline
        & \textit{AS-amd-100-bin} & 29.32\% \\
        \cite{he2018debin}-\texttt{amd64} & \textit{AS-i386-100-bin}\textcolor{blue}{$^{\mathrm{*}}$} & 13.84\% \\
        & \textit{AS-armel-100-bin}\textcolor{blue}{$^{\mathrm{*}}$} & 7.08\%\\
        & \textit{AS-mipsel-100-bin}\textcolor{blue}{$^{\mathrm{*}}$} & \textbf{-} \\
        \hline
        & \textit{AS-amd-100-bin}\textcolor{blue}{$^{\mathrm{*}}$} & 6.23\% \\
        \cite{he2018debin}-\texttt{i386} & \textit{AS-i386-100-bin} & 52.64\% \\
        & \textit{AS-armel-100-bin}\textcolor{blue}{$^{\mathrm{*}}$} & 1.05\%  \\
        & \textit{AS-mipsel-100-bin}\textcolor{blue}{$^{\mathrm{*}}$} & \textbf{-} \\
        \hline
        & \textit{AS-amd-100-bin}\textcolor{blue}{$^{\mathrm{*}}$} & 11.82\% \\
        \cite{he2018debin}-\texttt{armel} & \textit{AS-i386-100-bin}\textcolor{blue}{$^{\mathrm{*}}$} & 8.86\%  \\
        & \textit{AS-armel-100-bin} & 53.65\%  \\
        & \textit{AS-mipsel-100-bin}\textcolor{blue}{$^{\mathrm{*}}$} & \textbf{-} \\
        \hline
        \multicolumn{3}{l}{\textcolor{blue}{$^{\mathrm{*}}$} indicates that dataset was not used during the training.}\\
    \end{tabular}
    \vspace{-1.0em}
    \label{tab:crossarchi}
\end{table}

\subsection{The Practical Usage of CFG2VEC}
\label{sec:use_case}
In this section, we demonstrate how \textit{cfg2vec} assists REs in dealing with \textit{Defense Advanced Research Projects Agency} (DARPA) \textit{Assured MicroPatching} (AMP) challenges binaries.
The AMP program aims at enabling fast patching of legacy mission-critical system binaries, enhancing decompilation and guiding it toward a particular goal of a \textit{Reverse Engineer} (RE) by integrating the existing source code samples, the original build process information, and historical software artifacts.

\subsubsection{The MINDSIGHT project}  our multi-industry-academia initiative between Siemens, JHU/APL, BAE, and UCI jointly developed a project, \textit{Making Intelligible Decompiled Source by Imposing Homomorphic Transforms} (MINDSIGHT).
Our team focused on building an automated toolchain integrated with \textit{Ghidra}, aiming to enable the decompilation process with (1) a less granular identification of modular units, (2) an accurate reconstruction of symbol names, (3) the lifting of binaries to stylized C code, (4) a principled and scalable approach to reason about code similarity, and (5) the benchmarking of new decompilation techniques using state-of-the-art embedded software binary datasets. 
To date, our team has developed an open-source tool, \textit{CodeCut}\footnote{ \textit{CodeCut}'s repository: \url{https://github.com/DARPAMINDSIGHT/CodeCut}}, to improve the accuracy and completeness of \textit{Ghidra}’s module identification, providing an automated script-based decompilation analysis toolchain to ease the RE’s expert interpretation.
Besides, we also developed a \textit{Homomorphic Transform Language} (HTL) to describe transformations on \textit{Abstract Syntax Tree} (AST) languages and the rules of their composition.
Our tool, integrated with \textit{ghidra}, allows developers to transform the decompiled code syntactically while keeping it semantically equivalent.
The key idea is to use this HTL to morph a \textit{Ghidra} AST into a GCC AST to lift the decompiled binary to a high-level C representation. 
This process can make it easier for REs to comprehend the binary code.
\textit{cfg2vec} is another tool developed in the MINDSIGHT project, enabling the reconstruction of function names, saving the manual guesswork from REs.

\begin{figure}[!ht]
    \centering
    \includegraphics[width=\linewidth]{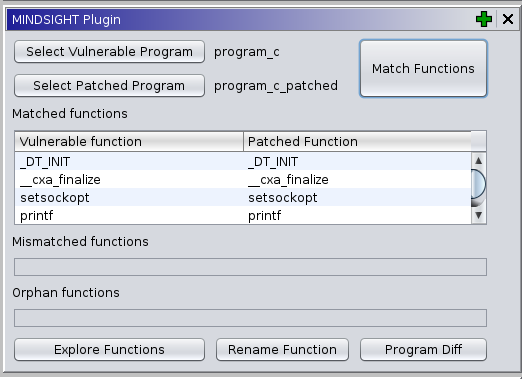}
    \vspace{-1em}
    \caption{The plugin screenshot integrated into Ghidra.}
    \vspace{-1.0em}
    \label{fig:plugin}
\end{figure}

\subsubsection{The cfg2vec plugin} In \textit{MINDSIGHT} project, we incorporated \textit{cfg2vec} into \textit{Ghidra} decompiler as a plugin application. 
Our \textit{cfg2vec} plugin assists REs in comprehending the binaries by providing a list of potential function names for each function without its name.
Technically, like all \textit{Ghidra} plugins, our \textit{cfg2vec} plugin bases on Java with its core inference modules implemented as a REST API in Python 3.8. 
Once the metadata of a stripped binary is extracted from \textit{Ghidra} decompiler, it is then sent to the \textit{cfg2vec} end-point, which calculates and returns the inferred mappings for all the functions.
Figure~\ref{fig:plugin} demonstrates the user interface of our \textit{cfg2vec} plugin.
In this scenario, the user must provide the vulnerable and the reference binary with extra debug information, such as function names. The ``Match Functions'' button triggers \textit{cfg2vec} functionality and displays the function mapping results in three tables:

\begin{itemize}
    \item \textit{Matched Table}: displays the mapping of similar functions.
    \item \textit{Mismatched Table}: displays the mapping of \textit{dissimilar} functions and, therefore, candidates for patching.
    \item \textit{Orphan Table}: displays the mapping of functions with a low confidence score.
\end{itemize}

The groupings reduce REs' workload.
Rather than inspecting all functions, they can focus on patching candidate functions (mismatched functions) and the orphans.
The ``Explore Functions'' button invokes Ghidra's function explorer, where the two functions can be compared side-by-side, as shown in Figure~\ref{fig:plugin}.
This utility allows the user to switch between C and assembly language, thus assisting in confirming or modifying the mappings from the three tables. 
Regarding \textit{cfg2vec}'s function prediction, the ``Rename Function'' button takes the selected row from the tables and imposes the name from the patched binary in the vulnerable binary.
When the ``Match Functions'' button fires, we invoke the FCG and CFG generators for the two programs (vulnerable and patched).

\subsubsection{The use-case for AMP challenge binaries} DARPA AMP challenges is about REs to patch a vulnerability regarding a weak encryption algorithm where the encryption of communication traffic was accomplished with a deprecated cipher suite, Triple DES or 3DES~\cite{callas_2017}.
For this challenge, REs have to analyze the vulnerable binary, identify functions and instructions to be patched, \textit{3DES cipher suite} in this case, and patch 3DES-related function calls and instructions with the ones for AES~\cite{bernstein_cobb_2021}. 
All these steps happen at the decompiled binary level, and the vulnerable binaries are optimized by a compiler and stripped of the debugging information and function names. 
Furthermore, these binaries are sometimes statically linked against libraries such as GNU C Library~\cite{rothwell2007gnu} or OpenSSL, which introduce many extra functions to the binary (some of which will never be called/used).
Given these complications, it becomes a non-trivial task for an RE to make sense of all these functions, find the problem, and successfully patch the problem. 
The direct usage of our \textit{cfg2vec} plugin was to pick a function of interest with stripped information and see predictions of potential function names or matching functions from the available reference binary to confirm that whether this function is in the critical path during RE's problem solving.
As Figure \ref{fig:plugin} shows, our plugin allows users to see possible matches between functions from a stripped vulnerable binary and functions from a patched (reference) binary with extra information.
REs may then leverage such information and make appropriate notes for that function, allowing them to complete their jobs more efficiently.
The main feedback we received from REs who used the tool was that this is the functionality REs would like to have.
However, the accuracy and usability of the tool were not high enough to truly utilize the tool's potential.

\section{Conclusion}
\label{sec:conclusion}
This paper presents \textit{cfg2vec}, a Hierarchical Graph Neural Network-based approach for software reverse engineering.
Building on top of \textit{Ghidra}, our \textit{cfg2vec} plugin can extract a \textit{Graph-of-Graph} (GoG) representation for binary, combining the information from Control-Flow Graphs (CFG) and Function-Call Graphs (FCG).
\textit{Cfg2vec} utilizes a hierarchical graph embedding framework to learn the representation for each function in binary code compiled into various architectures. 
Lastly, our \textit{cfg2vec} utilizes the learned function embeddings for function name prediction, outperforming the state-of-the-art \cite{he2018debin} by an average of 24.54\% across all tested binaries.
By increasing the amount of data, our model achieved 51.84\% better.
While \cite{he2018debin} requires training once for each CPU architecture, our \textit{cfg2vec} still can outperform consistently across all the architectures, only with one training.
Besides, our model generalizes the learning better \cite{he2018debin} to the binaries built from untrained CPU architectures.
Lastly, we demonstrate that our \textit{cfg2vec} can assist the real-world REs in resolving \textit{Darpa Assured MicroPatching} (AMP) challenges.

\ifCLASSOPTIONcompsoc
  \section*{Acknowledgments}
\else
  \section*{Acknowledgment}
\fi

This material is based upon work supported by the Defense Advanced Research Projects Agency (DARPA) and Naval Information Warfare Center Pacific (NIWC Pacific) under Contract Number N66001-20-C-4024. 
The views, opinions, and/or findings expressed are those of the author(s) and should not be interpreted as representing the official views or policies of the Department of Defense or the U.S. Government.

\ifCLASSOPTIONcaptionsoff
  \newpage
\fi

\bibliographystyle{IEEEtran}
\bibliography{IEEEabrv,cfg2vec}

 
\end{document}